\documentclass[
 reprint,
 amsmath,amssymb,
 aps,
 floatfix,
]{revtex4-2}

\usepackage{graphicx}
\usepackage{bm}
\usepackage[colorlinks=true,citecolor=blue,urlcolor=blue,linkcolor=blue]{hyperref}

\begin{document}

\title{Exceptional Cones from an Indefinite Bogoliubov Metric in Hyperbolic Polariton Condensates}

\author{Junhui Cao$^{1}$}%
\email{tsao.c@mipt.ru}
\author{Kirill Bazarov$^{1}$}%
\author{Alexey Kavokin$^{1,2,3}$}%
\email{a.kavokin@westlake.edu.cn}
\affiliation{$^{1}$Abrikosov Center for Theoretical Physics, Moscow Center for Advanced Studies, Moscow 141701, Russia\\
$^{2}$Department of Physics, St. Petersburg State University, University Embankment, 7/9, St. Petersburg, 199034, Russia\\
$^{3}$Russian Quantum Center, 30, Bolshoy Boulevard, Skolkovo, Russia
}

\date{\today}

\begin{abstract}
Long-wavelength Bogoliubov phonons in an ordinary condensate realize the standard acoustic Lorentz metric. We show that a condensate formed in a hyperbolic polariton band realizes a different collective geometry with an indefinite Bogoliubov metric whose spatial signature is inherited from the opposite signs of the band curvatures. This metric converts the acoustic light cone into a hyperbolic stability wedge, separating propagating quasiparticles from dynamically unstable ones. In a driven-dissipative condensate, gain saturation turns this metric relation into a non-Hermitian Bogoliubov dispersion. The zero-discriminant surface becomes an exceptional cone in the parametric space $(q_x,q_y,\Delta_{\rm NH})$, appearing as an exceptional hyperbola at fixed gain saturation. Across this surface the Bogoliubov branches coalesce, the biorthogonal phase rigidity collapses, and the spectrum changes from propagation to amplified or overdamped dynamics. Our results identify hyperbolic polariton condensates as a controllable setting where non-Hermitian exceptional degeneracies are organized by an effective Bogoliubov metric.
\end{abstract}

\maketitle

\paragraph*{Introduction.---}
Analogue gravity usually begins with a positive answer to a kinematic question: can excitations of a fluid behave as fields on an effective spacetime? In a weakly interacting Bose condensate the long-wavelength phase mode obeys an acoustic relativistic wave equation, with the sound velocity and the background flow defining the effective Lorentz metric~\cite{unruh1981,visser1998, garay2000,barcelo2003towards,recati2009bogoliubov}. Exciton polaritons are a natural platform for such physics because their dispersion, interactions, gain, loss, and flow can be engineered in microcavities and photonic lattices~\cite{carusotto2013, byrnes2014,bloch2022,nguyen2015,wouters2007,jia2025femtosecond,cao2025acoustic}.

Hyperbolic polaritonic media introduce a second geometric ingredient. Opposite signs of the effective masses generate open isofrequency contours and directional propagation in hyperbolic metamaterials and van der Waals polaritonics~\cite{poddubny2013,basov2016,low2017,sedov2015hyperbolic,wang2024planar,yoxall2015direct}. However, the metric commonly used there is a passive optical metric. Once the mode is macroscopically occupied, the low-energy excitations are Bogoliubov quasiparticles, controlled by single-particle dispersion curvature, self-interactions, condensate density, particle-hole mixing, and non-Hermitian gain saturation. Hence, the central question is what geometry is seen by collective Bogoliubov excitations when the condensate is formed in a hyperbolic band.

Our answer has three parts. First, the conservative long-wavelength Bogoliubov metric is spatially indefinite. Second, driven-dissipative gain saturation does not simply add a linewidth. It also promotes the metric relation to a driven-dissipative Bogoliubov dispersion. Third, the exceptional degeneracy of the non-Hermitian Bogoliubov spectrum is the zero-discriminant surface associated with the same indefinite quadratic form. The resulting object is an exceptional cone in the enlarged parametric space $(q_x,q_y,\Delta_{\rm NH})$ and an exceptional hyperbola in the momentum plane of fixed gain saturation. This distinction matters because exceptional points in non-Hermitian photonics and condensed-matter systems are often introduced as local degeneracies of a few-mode Hamiltonian~\cite{elganainy2018,miri2019,ashida2020,bergholtz2021,PhysRevLett.120.065301}. Here the degeneracy is tied to a continuum collective metric. The same quadratic form that defines the long-wavelength stability sectors also determines the zero-discriminant manifold where the Bogoliubov eigenvectors become self-orthogonal.

\paragraph*{Model.---}
We use dimensionless units and consider a scalar driven-dissipative condensate \cite{PhysRevLett.100.250401,wouters2008spatial},
\begin{equation}
i\partial_t\psi=\left[\epsilon(-i\nabla)+g|\psi|^2+\frac{i}{2}\left(P-\gamma-\eta|\psi|^2\right)\right]\psi ,
\label{eq:gpe}
\end{equation}
with hyperbolic quadratic dispersion
\begin{equation}
\epsilon_{\mathbf q}=\alpha_xq_x^2+\alpha_yq_y^2,\qquad \alpha_x\alpha_y<0 .
\label{eq:single_particle}
\end{equation}
Here $\psi(\mathbf r,t)$ is the condensate order parameter. The nonlinear term $g|\psi|^2$ describes the mean-field polariton-polariton interaction. For $g>0$ it gives the usual density-dependent blueshift of the condensate energy. The last term is purely imaginary and accounts for the driven-dissipative character of the polariton condensate. The parameter $P$ is the linear gain supplied by the nonresonant pump, $\gamma$ is the linear polariton loss rate, and $\eta|\psi|^2$ describes nonlinear gain saturation.

The uniform state has $n_0=(P-\gamma)/\eta$ and $\mu=gn_0$. Linearizing about
$\psi_0=\sqrt{n_0}e^{-i\mu t}$ gives the non-Hermitian Bogoliubov matrix
\begin{equation}
\mathcal L(\mathbf q)=
\begin{pmatrix}
\epsilon_{\mathbf q}+A & A\\
-A^* & -\epsilon_{\mathbf q}-A^*
\end{pmatrix},
\qquad
A=gn_0-i\Delta_{\rm NH},
\label{eq:bdg_matrix}
\end{equation}
where $\Delta_{\rm NH}=\eta n_0/2$. Its eigenfrequencies are
\begin{equation}
\omega_\pm(\mathbf q)=
-i\Delta_{\rm NH}\pm
\sqrt{\epsilon_{\mathbf q}\left(\epsilon_{\mathbf q}+2gn_0\right)
-\Delta_{\rm NH}^2}.
\label{eq:exact_spectrum}
\end{equation}
Information about the conservative metric, the stability sectors, and the exceptional condition is all contained in the radicand of $\omega_\pm(\mathbf{q})$,
\begin{equation}
D(\mathbf q)=
\epsilon_{\mathbf q}\left(\epsilon_{\mathbf q}+2gn_0\right)-\Delta_{\rm NH}^2 .
\label{eq:discriminant}
\end{equation}
For positive effective masses, Eq.~\eqref{eq:exact_spectrum} reduces to the familiar driven-dissipative Bogoliubov structure of a polariton condensate, including a Goldstone branch and an amplitude-relaxation branch ~\cite{wouters2007}. In a hyperbolic band, the opposite curvatures change the spatial signature of the Bogoliubov metric. As a consequence, the isotropic exceptional ring is deformed into a hyperbolic exceptional manifold, and amplification becomes locked to specific angular sectors in momentum space.

\begin{figure}[t]
    \centering
    \includegraphics[width=\linewidth]{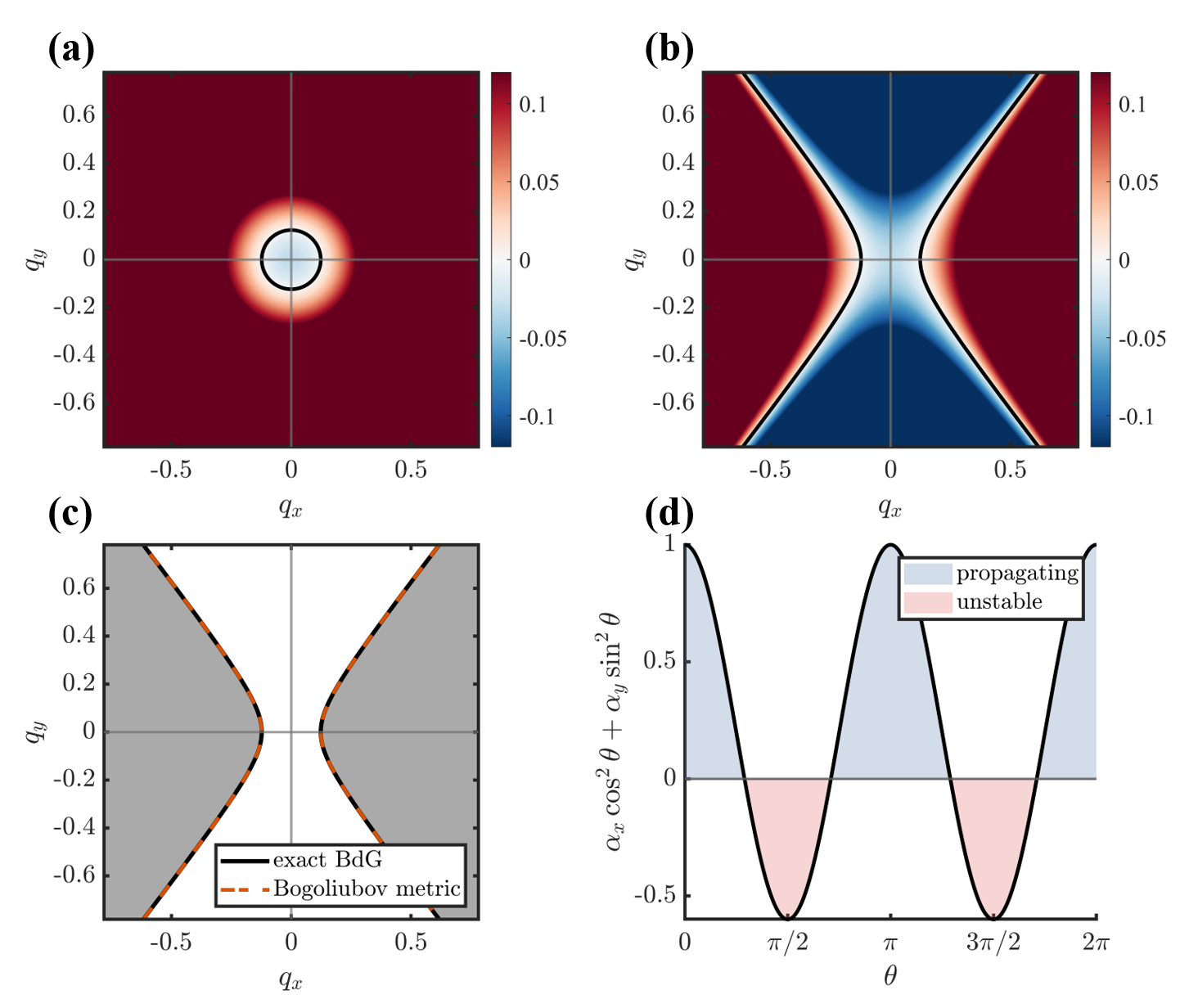}
    \caption{\label{fig:metric_topology} 
    Metric origin of the exceptional contour. (a) For an ordinary condensate with $\alpha_x=\alpha_y=1$, the exact non-Hermitian condition $D(\mathbf q)=0$ gives a closed exceptional ring. (b) For a hyperbolic condensate with $\alpha_x=1,\alpha_y=-0.6$, the same condition opens into a hyperbola. (c) The exact contour is almost indistinguishable from the Bogoliubov metric prediction close to the origin. For $\eta n_0=0.35$ the $q_y=0$ exceptional momentum differs by only $0.38\%$. (d) The sign of $\alpha_x\cos^2\theta+\alpha_y\sin^2\theta$ selects propagating and dynamically unstable angular regimes.}
\end{figure}

Figure~\ref{fig:metric_topology} shows how the fixed-gain section changes topology from an ordinary ring (Fig.~\ref{fig:metric_topology}a) to a hyperbola (Fig.~\ref{fig:metric_topology}b) when the curvature signature changes. The full BdG contour (Fig.~\ref{fig:metric_topology}c) is obtained from $D(\mathbf q)=0$, while the long-wavelength contour is obtained by keeping only $2gn_0\epsilon_{\mathbf q}-\Delta_{\rm NH}^2$. The two agree near the origin. For $gn_0=1$, $\alpha_x=1$, $\alpha_y=-0.6$, and $\eta n_0=0.35$, the exact low-energy root is $q_{\rm EP}=0.123276$, while the metric formula gives $0.123744$. This shows that the exceptional degeneracy is not an accidental two-mode crossing. It is the zero-discriminant surface associated with the same indefinite quadratic form that controls the long-wavelength Bogoliubov stability.

Set $\Delta_{\rm NH}=0$. The conservative BdG dispersion is
\begin{equation}
\omega_\pm^2=\epsilon_{\mathbf q}\left(\epsilon_{\mathbf q}+2gn_0\right).
\label{eq:conservative_bdg}
\end{equation}
In the long-wavelength limit this reduces to
\begin{equation}
\omega^2\approx 2gn_0(\alpha_xq_x^2+\alpha_yq_y^2).
\label{eq:hydro_general}
\end{equation}
For positive curvatures, Eq.~\eqref{eq:hydro_general} is the usual acoustic metric relation. For a hyperbolic condensate with $\alpha_x>0$ and $\alpha_y<0$, we define $c_x^2=2gn_0\alpha_x$ and $c_y^2=2gn_0|\alpha_y|$. Then
\begin{equation}
\omega^2-c_x^2q_x^2+c_y^2q_y^2=0 .
\label{eq:indefinite_metric}
\end{equation}
Equivalently, this equation can be written as a null condition for an effective Bogoliubov metric,
\begin{equation}
    g_{\rm B}^{\mu\nu}q_\mu q_\nu=0,
    \qquad q_\mu=(\omega,q_x,q_y),
    \label{eq:metric_relation}
\end{equation}
Here $g_{\rm B}^{\mu\nu}$ is the inverse effective metric that organizes the long-wavelength propagation of Bogoliubov phase excitations, in the same spirit as acoustic gravity. Up to an overall conformal factor, it is given by
\begin{equation}
    g_{\rm B}^{\mu\nu}=
    \begin{pmatrix}
    1 & 0 & 0\\
    0 & -c_x^2 & 0\\
    0 & 0 & c_y^2
    \end{pmatrix}.
    \label{eq:metric_tensor}
\end{equation}
The opposite signs of the two spatial entries are the origin of both the hyperbolic causal wedge and the direction-dependent Bogoliubov stability. This sign structure is the key feature: the spatial part of the Bogoliubov metric is indefinite. It follows that
\begin{equation}
c_x^2q_x^2>c_y^2q_y^2\Rightarrow\omega\in\mathbb R,\qquad
c_x^2q_x^2<c_y^2q_y^2\Rightarrow\omega\in i\mathbb R .
\label{eq:stability_wedge}
\end{equation}
The metric controls both propagation and dynamical stability. The hyperbolic band gives rise to anisotropic quasiparticle propagation together with a sector-dependent stability structure in momentum space, with qualitatively different time evolution.

\begin{figure}[t]
    \centering
    \includegraphics[width=\linewidth]{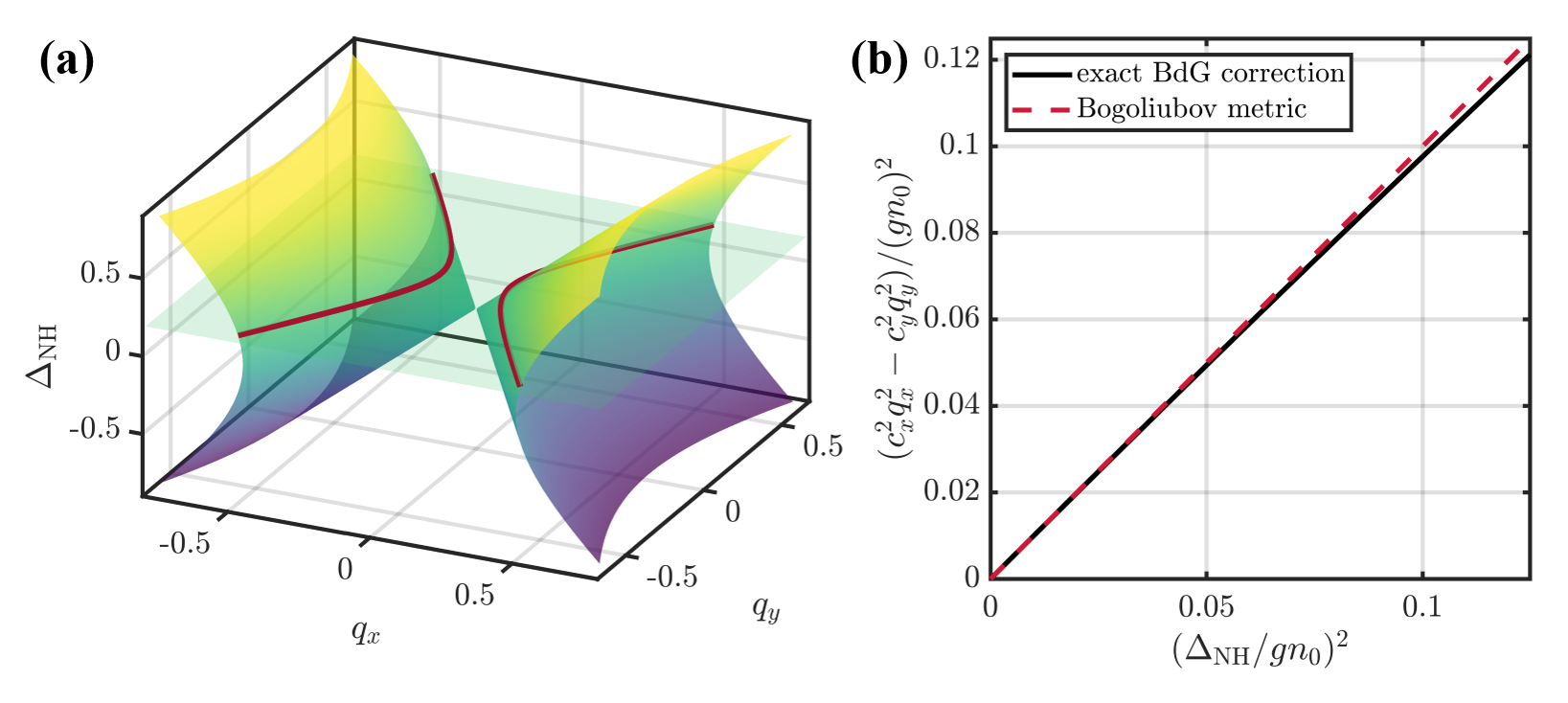}
\caption{\label{fig:augmented_cone}
    Exceptional cone and metric collapse. (a) The zero-discriminant condition $c_x^2q_x^2-c_y^2q_y^2=\Delta_{\rm NH}^2$ forms a cone in the augmented space $(q_x,q_y,\Delta_{\rm NH})$. The red plane marks a fixed gain saturation. (b) Exact BdG exceptional points, sampled over several gain saturations and hyperbolic curvatures, collapse when plotted as $(c_x^2q_x^2-c_y^2q_y^2)/(gn_0)^2$ versus $(\Delta_{\rm NH}/gn_0)^2$. The dashed diagonal is the long-wavelength Bogoliubov metric prediction, while the solid curve is the finite-momentum BdG correction $2[\sqrt{1+(\Delta_{\rm NH}/gn_0)^2}-1]$.}
\end{figure}

\paragraph*{Exceptional cone.---}
We define the centered frequency
\begin{equation}
\Omega=\omega+i\Delta_{\rm NH}.
\label{eq:centered_frequency}
\end{equation}
With gain saturation, the small-$q$ form of Eq.~\eqref{eq:exact_spectrum} becomes the driven-dissipative Bogoliubov dispersion relation
\begin{equation}
\Omega^2-c_x^2q_x^2+c_y^2q_y^2+\Delta_{\rm NH}^2=0 .
\label{eq:complexified_mass_shell}
\end{equation}
Formally, Eq.~\eqref{eq:complexified_mass_shell} resembles a relativistic energy-momentum relation,
\begin{equation}
E^2-c^2p^2-m^2c^4=0 ,
\end{equation}
but with two important differences. First, the spatial part is hyperbolic, because the $q_x$ and $q_y$ terms have opposite signs. Second, the last term is not a rest-mass energy, but a non-Hermitian scale set by gain saturation. The exceptional degeneracy occurs when the square-root splitting vanishes, or equivalently when $\Omega=0$ on this long-wavelength dispersion relation. This gives
\begin{equation}
c_x^2q_x^2-c_y^2q_y^2=\Delta_{\rm NH}^2 .
\label{eq:exceptional_cone}
\end{equation}
The geometry of Eq.~\eqref{eq:exceptional_cone} is shown in Fig.~\ref{fig:augmented_cone}. In the enlarged space $(q_x,q_y,\Delta_{\rm NH})$, the long-wavelength zero-discriminant condition forms an exceptional cone. Fixing the gain saturation selects a horizontal section of this surface, which is measured as an exceptional hyperbola in the two-dimensional momentum plane. The dashed metric prediction and the solid exact BdG contour in Fig.~\ref{fig:augmented_cone}(b) coincide near the origin, confirming that the exceptional manifold is organized by the same long-wavelength Bogoliubov metric that controls the stability sectors.

\begin{figure}[t]
    \centering
    \includegraphics[width=\linewidth]{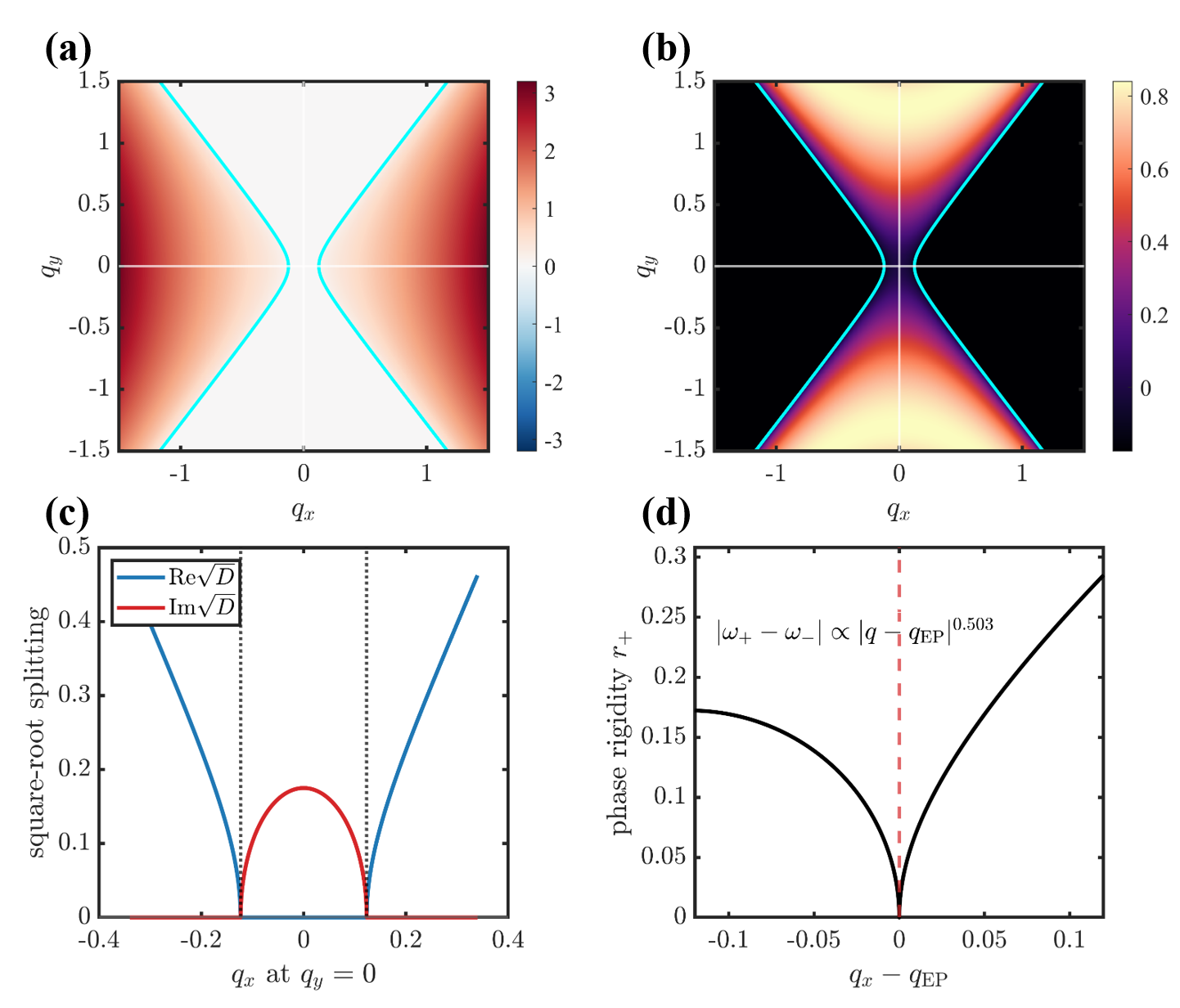}
    \caption{\label{fig:ep_diagnostics}
    Spectral and eigenvector diagnostics of the exceptional hyperbola. (a,b) Maps of the real and imaginary parts of $\omega_+$, with the cyan line marking $D(\mathbf q)=0$. (c) Along $q_y=0$, the centered splitting $\sqrt{D}$ changes from real to imaginary across the exceptional points. (d) A log-log fit near the right exceptional point gives $|\omega_+-\omega_-|\propto |q-q_{\rm EP}|^{0.503}$, while the biorthogonal phase rigidity of the positive branch collapses to $r_+= 1.2\times 10^{-8}$ at the EP, which demonstrates a defective non-Hermitian degeneracy.}
\end{figure}

The condition $D(\mathbf q)=0$ gives the coalescence of the two eigenvalues, but a non-Hermitian exceptional point also requires the eigenvectors to coalesce. We verified this directly from Eq.~\eqref{eq:bdg_matrix}. Along the $q_y=0$ cut, the two exceptional points occur at $q_x=\pm q_{\rm EP}$, where
\begin{equation}
q_{\rm EP}^2=-gn_0+\sqrt{g^2n_0^2+\Delta_{\rm NH}^2}.
\label{eq:qep_exact}
\end{equation}
Near either point the discriminant is linear in $q_x-q_{\rm EP}$, so the branch splitting must scale as a square root. Numerically, fitting $|\omega_+-\omega_-|$ over more than three decades in $|q_x-q_{\rm EP}|$ gives an exponent $0.503$. In addition, the biorthogonal phase rigidity
\begin{equation}
r_+=\frac{|\langle L_+|R_+\rangle|}{\sqrt{\langle L_+|L_+\rangle\langle R_+|R_+\rangle}}
\label{eq:phase_rigidity}
\end{equation}
collapses at the exceptional point. Both features are shown in Figure~\ref{fig:ep_diagnostics}d.

\begin{figure}[t]
    \centering
    \includegraphics[width=\linewidth]{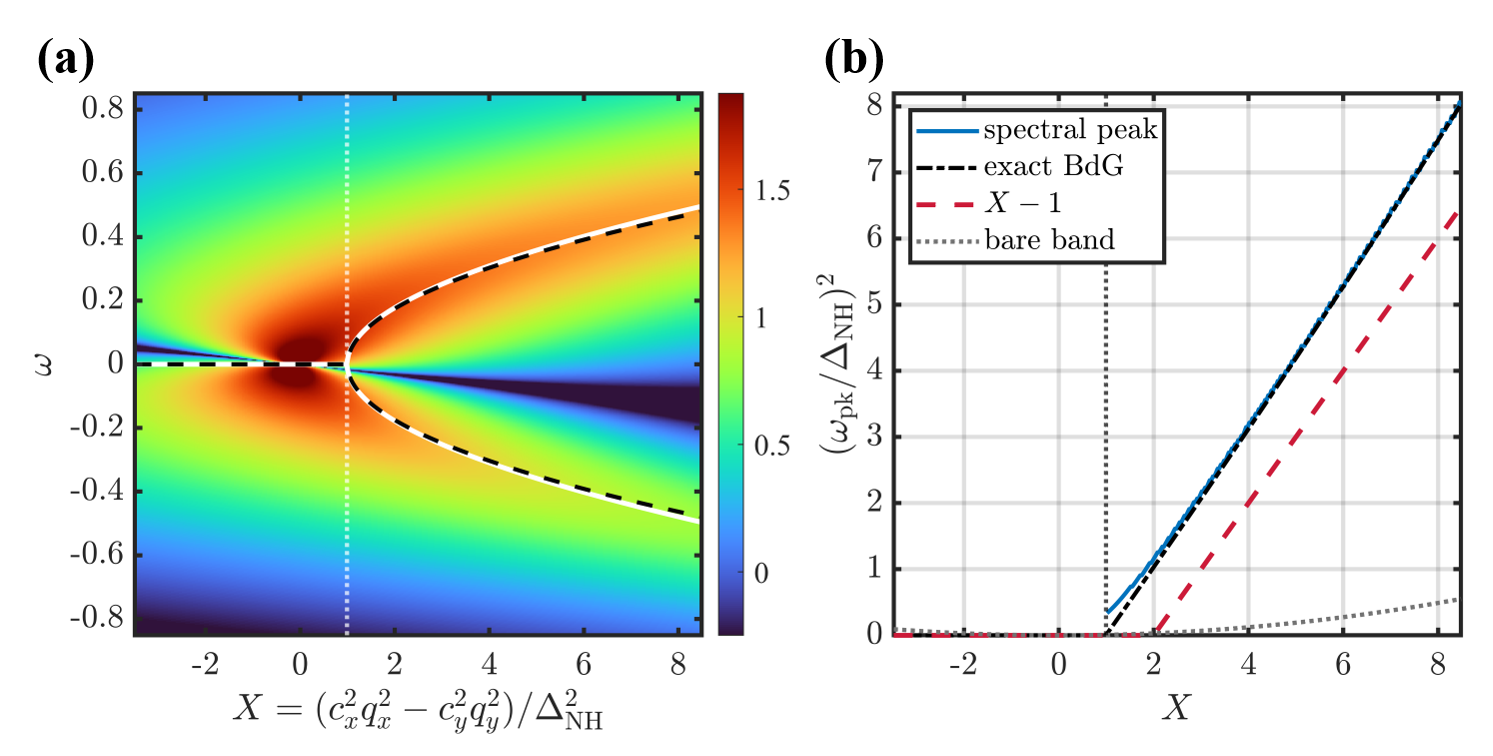}
    \caption{\label{fig:spectral_observable}
    Metric-resolved spectral observable. (a) Particle spectral function $|-2\,\mathrm{Im}\,G^R_{11}|$ after projecting momenta onto $X=(c_x^2q_x^2-c_y^2q_y^2)/\Delta_{\rm NH}^2$. The solid line is the exact BdG branch, the dashed line is the Bogoliubov metric limit, and the vertical dotted line marks $X_{\rm EP}$. (b) The positive spectral peak in the propagating sector collapses onto the exact BdG hyperboloid, while the bare single-particle dispersion fails. For the representative GaAs polariton parameters \cite{tsintzos2008gaas,sun2017direct}, $m_x=9.0\times10^{-5}m_e$, $m_y=-1.5\times10^{-4}m_e$, and $gn_0=0.50~{\rm meV}$, the coefficients are $\alpha_x=0.423~{\rm meV\,\mu m^2}$ and $\alpha_y=-0.254~{\rm meV\,\mu m^2}$. The parameters used here correspond to $\Delta_{\rm NH}=0.0875~{\rm meV}$, $q_{\rm EP}=0.134~\mu{\rm m}^{-1}$ along $q_y=0$, and a characteristic time $\hbar/\Delta_{\rm NH}=7.52~{\rm ps}$.}
\end{figure}

One decisive test of the Bogoliubov metric is to collapse the spectrum onto the metric coordinate
\begin{equation}
X=\frac{c_x^2q_x^2-c_y^2q_y^2}{\Delta_{\rm NH}^2}= \frac{2gn_0\epsilon_{\mathbf q}}{\Delta_{\rm NH}^2}.
\label{eq:metric_coordinate}
\end{equation}
Angle-resolved spectroscopy \cite{stepanov2019dispersion,claude2021high} is governed by the
retarded Green function
\begin{equation}
G^R(\mathbf q,\omega)=\left[(\omega+i\kappa)\mathbb{I}-L(\mathbf q)\right]^{-1},
\label{eq:retarded_green}
\end{equation}
where $\kappa$ is a small broadening. We define the metric-resolved particle spectral function by binning $A_{11}(\mathbf q,\omega)=-2\,{\rm Im}\,G^R_{11}(\mathbf q,\omega)$ at fixed $X$. In the propagating regime, the spectral ridge must obey the exact BdG relation
\begin{equation}
\left(\frac{\omega_{\rm pk}}{\Delta_{\rm NH}}\right)^2=X-1+\frac{\Delta_{\rm NH}^2}{4g^2n_0^2}X^2.
\label{eq:spectral_mass_shell}
\end{equation}
This differs from the bare single-particle dispersion $|\epsilon_{\mathbf q}|=\Delta_{\rm NH}^2|X|/(2gn_0)$. Figure~\ref{fig:spectral_observable} shows the result. The spectral ridge collapses onto the exact BdG branch for $X>X_{\rm EP}=0.99246$, while the anomalous fraction
\begin{equation}
f_{12}=\frac{|G^R_{12}|^2}{|G^R_{11}|^2+|G^R_{12}|^2}
\label{eq:anomalous_fraction}
\end{equation}
remains sizable near the threshold, with a mean value around 0.44, showing that the observed response follows the BdG dispersion while retaining strong particle-hole mixing.

\begin{figure}[t]
    \centering
    \includegraphics[width=\linewidth]{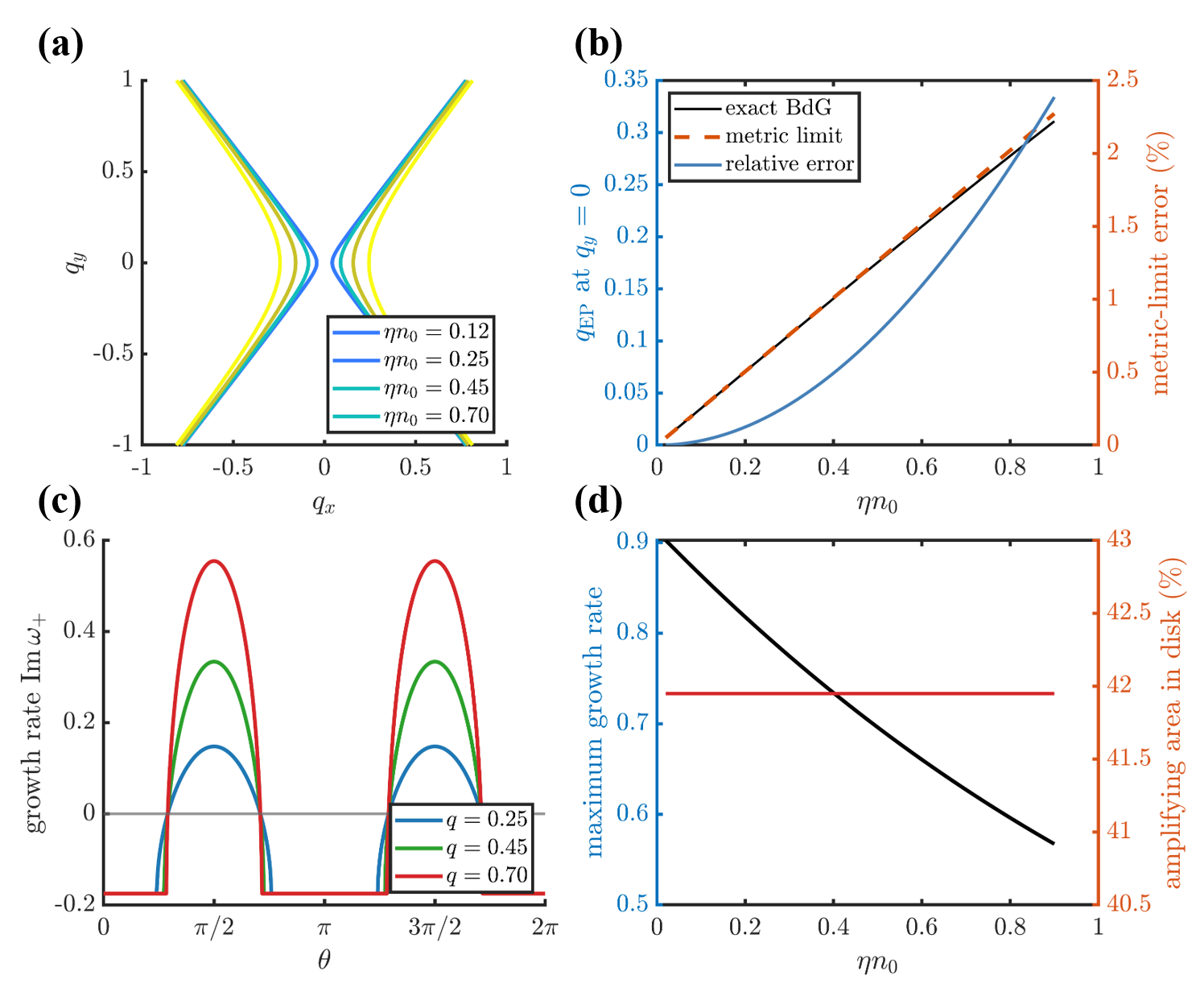}
    \caption{\label{fig:tunability}
    Tunability and angular selectivity. (a) Increasing $\eta n_0$ moves the exceptional hyperbola outward, making the non-Hermitian null surface directly controllable by gain saturation and density. (b) The exact $q_{\rm EP}$ follows the Bogoliubov metric prediction at small $\eta n_0$. The relative error remains below a few percent over the range shown. (c) The growth rate $\mathrm{Im}\,\omega_+$ is strongly angle-dependent where only momenta in the hyperbolic unstable sectors are amplified. (d) Loss reduces the maximum growth rate, while the fraction of amplifying momenta inside the unit disk remains approximately fixed by the metric.}
\end{figure}

The exceptional cone, and hence its fixed-gain exceptional hyperbola, is also tunable. Equation~\eqref{eq:exceptional_cone} shows that increasing $\Delta_{\rm NH}$ pushes the null surface outward. The exact formula Eq.~\eqref{eq:qep_exact} gives the same behavior beyond the long-wavelength approximation. Figure~\ref{fig:tunability}(a,b) compares the exact BdG contour with the metric scaling as $\eta n_0$ is varied. At the baseline value
$\eta n_0=0.35$, the long-wavelength error in $q_{\rm EP}$ is only $0.38\%$. The same calculation makes a sharper prediction for time-domain experiments, where amplification is direction selective. At fixed $|\mathbf q|$, the sign of the metric form controls whether the positive branch has a positive imaginary part. For the parameters of Fig.~\ref{fig:tunability}, the maximum growth rate inside a unit momentum disk is $\mathrm{Im}\,\omega_+\approx 0.754$, while about $42\%$ of momenta in the disk are amplifying. Increasing loss will reduce the maximum growth, but it does not erase the angular wedge geometry because the wedge is set primarily by the dispersion curvature signature.

\begin{figure}[t]
    \centering
    \includegraphics[width=\linewidth]{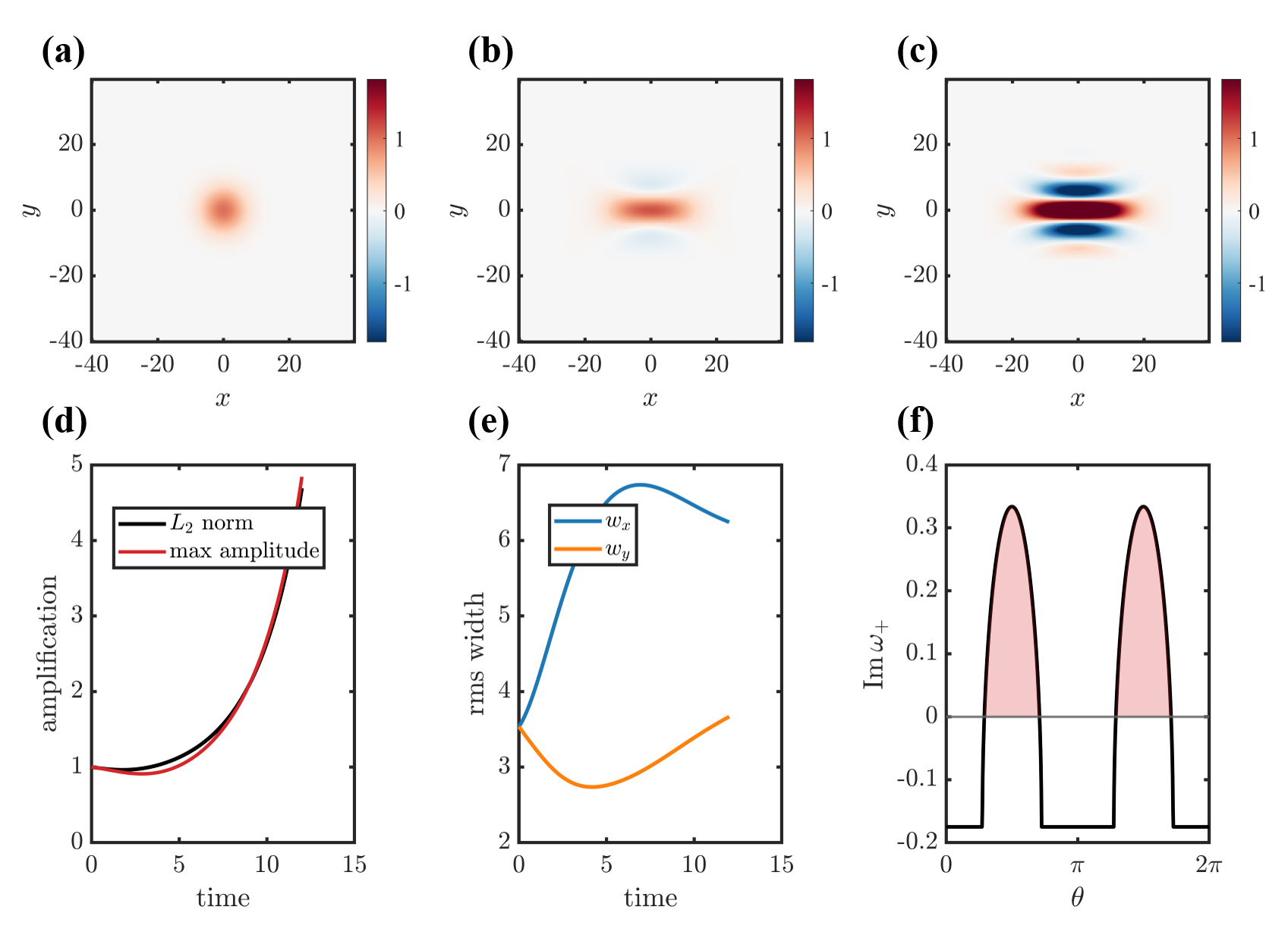}
    \caption{\label{fig:wavepacket}
    Real-space consequence of the hyperbolic Bogoliubov metric. (a-c) A localized phase-like perturbation propagated with $\delta\theta(\mathbf q,t)=\delta\theta(\mathbf q,0)e^{-i\omega_+t}$ becomes elongated and sign-alternating because stable and unstable angular sectors evolve differently, at $t=0,6,12$, respectively. (d) The $L_2$ norm and maximum amplitude grow by factors of $4.69$ and $4.84$ by $t=12$. (e) The root mean square widths become anisotropic, with $w_x=6.24$ and $w_y=3.67$ at $t=12$. (f) The angular gain profile. Two momentum lobes are amplified. }
\end{figure}

To connect the fixed-gain exceptional hyperbola to a real-space observable, we propagate a localized phase-like perturbation under the positive branch,
\begin{equation}
\delta\theta(\mathbf q,t)=\delta\theta(\mathbf q,0) \exp[-i\omega_+(\mathbf q)t].
\label{eq:wavepacket}
\end{equation}
Here we use the linear propagation to identify the momentum sectors and real-space directions selected by the Bogoliubov spectrum. As shown in Fig.~\ref{fig:wavepacket}(a-c), a circular Gaussian develops an elongated, sign-changing profile aligned with the directions selected by the unstable hyperbolic sectors. The integrated norm and maximum amplitude grow by nearly five times at $t=12$ in Fig.~\ref{fig:wavepacket}d, and the root mean square widths become strongly anisotropic (Fig.~\ref{fig:wavepacket}e). These signatures should be accessible in time-resolved pump-probe measurements of polariton condensates, while the real and imaginary parts of Eq.~\eqref{eq:exact_spectrum} can be reconstructed from angle-resolved photoluminescence and linewidths.

\paragraph*{Conclusion.---}
The physical distinction between an ordinary condensate and a hyperbolic condensate is more than quantitative anisotropy. In an ordinary condensate, Bogoliubov phonons see an acoustic Lorentz metric with a positive spatial quadratic form. In a hyperbolic polariton condensate, the collective particle-hole mode inherits the opposite signs of the effective masses and therefore sees an indefinite Bogoliubov metric. Driven-dissipative gain saturation then modifies the corresponding metric relation, whose zero-discriminant surface becomes an exceptional cone. This interpretation unifies the long-wavelength stability wedge geometry, the topology of the exceptional hyperbola, the square-root spectral splitting, eigenvector self-orthogonality, and the directional amplification of wave packets. The metric describes an effective geometry for collective Bogoliubov excitations, in the same spirit as acoustic gravity. The new element is the combination of an indefinite spatial signature, inherited from the hyperbolic band, with a non-Hermitian zero-discriminant condition generated by gain saturation. This combination turns isolated exceptional-point physics into an extended exceptional null surface.

Experimentally, the required hyperbolic band curvature may be realized near a saddle point of an engineered polariton lattice, a photonic-crystal polariton band, or a strongly anisotropic microcavity mode. The condensate does not need to be the thermodynamic ground state. In a driven polariton system, the occupied momentum state is selected by gain, pump geometry, relaxation, and nonlinear saturation. The unstable sectors discussed above are the measurable linear response of a gain-selected hyperbolic condensate. Because polariton platforms allow control over band curvature, density, pump power, and gain saturation, the exceptional cone can in principle be opened, displaced, and measured.

\begin{acknowledgements}
AK acknowledges support from Saint Petersburg State University (Grant No. 125022803069-4). 
\end{acknowledgements}

\bibliography{apssamp}

\end{document}